\documentclass[sigconf]{acmart}

\begin{document}
	
\author{Elias Kuiter}
\affiliation{
	\institution{Otto-von-Guericke University}
	\city{Magdeburg}
	\country{Germany}
}
\email{kuiter@ovgu.de}
\author{Gunter Saake}
\affiliation{
	\institution{Otto-von-Guericke University}
	\city{Magdeburg}
	\country{Germany}
}
\email{saake@ovgu.de}
\renewcommand{\shortauthors}{E. Kuiter}
\title{A Survey and Comparison of Industrial and Academic Research on the Evolution of Software Product Lines}

\begin{abstract}
	Past research on software product lines has focused on the initial development of reusable assets and related challenges, such as cost estimation and implementation issues.
	Naturally, as software product lines are increasingly adopted throughout industry, their ongoing maintenance and evolution are getting more attention as well.
	However, it is not clear to what degree research is following this trend, and where the interests and demands of the industry lie.
	In this technical report, we provide a survey and comparison of selected publications on software product line maintenance and evolution at SPLC.
	In particular, we analyze and discuss similarities and differences of these papers with regard to their affiliation with industry and academia.
	From this, we infer directions for future research that pave the way for systematic and organized evolution of software product lines, from which industry may benefit as well.
\end{abstract}

\keywords{Software Product Line, Evolution, Industry, Academia, SPLC}

\ccsdesc[500]{Software and its engineering~Software product lines}
\ccsdesc[300]{Software and its engineering~Software evolution}
\ccsdesc[300]{Software and its engineering~Maintaining software}

\settopmatter{printacmref=false}
\setcopyright{none}
\renewcommand\footnotetextcopyrightpermission[1]{}
\pagestyle{plain}

\maketitle

\section{Introduction}
\label{sec:introduction}

Customizable and highly-configurable software systems play a major role in today's software landscape~\cite{Fischer2014, Pohl2005}.
\emph{Software product lines} (SPLs) are a systematic approach to create and maintain such highly-configurable software systems.
With SPLs, customized variants of software systems can be compiled from reusable artifacts, according to the customer's particular requirements.
Thus, SPLs allow mass customization, while still promising reduced development costs for software vendors after initial investments~\cite{Clements2002, Knauber2001, Pohl2005}.

Past research on SPLs has mainly been motivated by industrial needs, while industry has also benefited from developments in research, such as variability modeling techniques~\cite{Berger2013, Rabiser2018}.
For example, it has been well-investigated how companies can transition to an SPL-based approach by extracting an SPL from a monolithic legacy software system~\cite{Kruger2016, Krueger2002, Clements2002a}.

However, the increasing adoption of SPL methodologies in industry introduces new challenges.
In particular, the question arises how an organization may maintain and evolve an SPL after it has been established with one of the aforementioned techniques~\cite{McGregor2003}.
This is particularly relevant because SPLs require increased upfront investments and are intended as long-lasting assets to an organization.
Consequently, the relative importance of SPL \emph{maintenance} and \emph{evolution} in industry as well as research has grown with the increasing maturity of the SPL community.

\begin{figure}
	\vspace{4ex}
	\includegraphics[width=\linewidth]{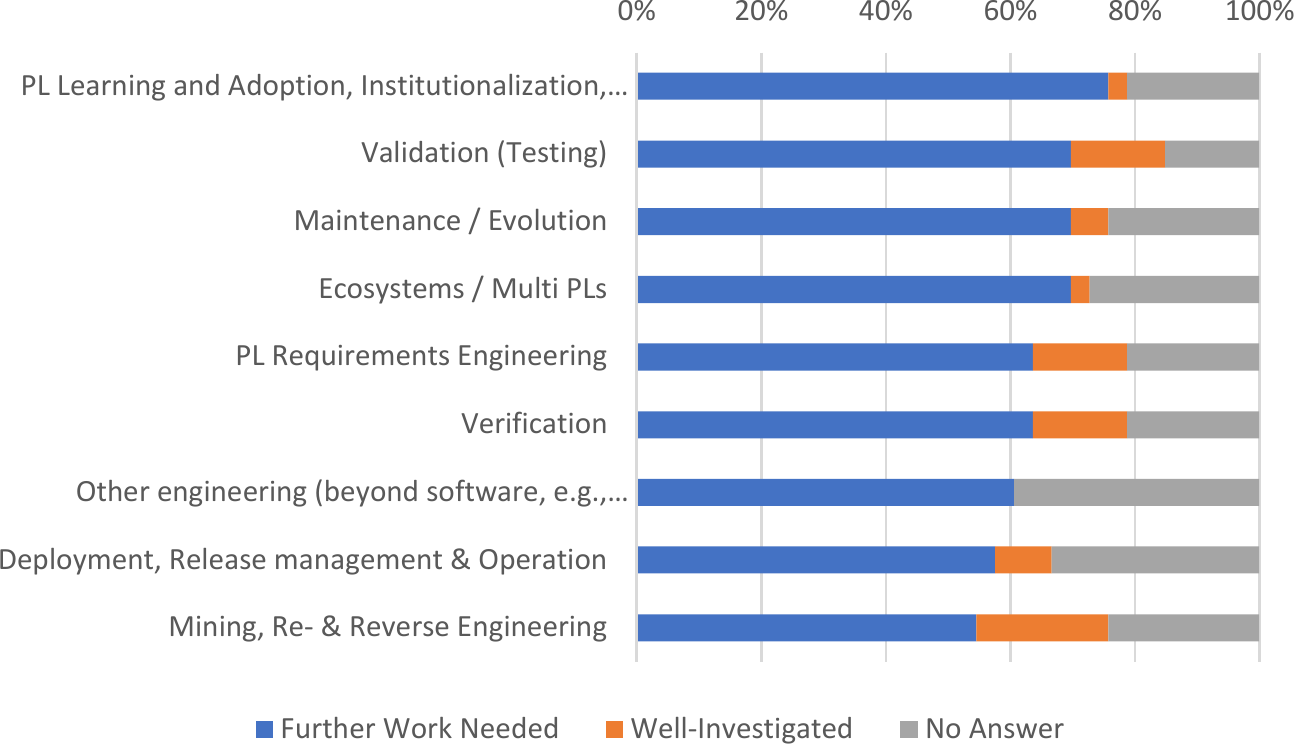}
	\caption{Top 10 of 27 research interests among 33 SPL researchers and practitioners as of 2018~\cite{RabiserMaterial}.}
	\label{fig:interests}
\end{figure}

In fact, evolution is currently one of the largest research interests in the SPL community:
In~\autoref{fig:interests}, we show a ranking of the most demanded activities in SPL engineering, which is based on a yet unpublished survey carried out at SPLC 2018 by~\citet{RabiserMaterial}.
In this survey, 33 SPL researchers and practitioners, having on average eight years of experience in this field, rated 27 different activities according to whether they were already well-investigated or needed further work in the future.
Using this information, we rank the activities according to the participants' opinions and show the ten activities with most interest in \autoref{fig:interests}.
In particular, 70\% of the participants think that SPL maintenance and evolution warrants further investigation, on par with SPL validation and ecosystems, and only 6\% of the participants judge it as well-investigated.

In this technical report, we survey and discuss selected papers in detail that address evolution of SPLs and have been published at SPLC.
In particular, we contribute the following:

\begin{itemize}
	\item We highlight commonalities, differences, and insights regarding SPL evolution from the selected papers.
	\item We distinguish between works from academia and industry, compare them, and give directions for future research.
\end{itemize}

Extending the work of \citet{Rabiser2018}, we provide deeper insights regarding product-line research on maintenance and evolution by taking a closer look at actual publications on this topic.

The remainder of this report is organized as follows:
In \autoref{sec:background}, we give background regarding SPLs and their evolution.
In \autoref{sec:method}, we describe the methodology of our survey.
In \autoref{sec:results}, we present the results of our survey, which we discuss and compare in \autoref{sec:discussion}.
We further discuss threats to validity in \autoref{sec:validity} and conclude the report in \autoref{sec:conclusion}.

\section{Background}
\label{sec:background}

We start by giving a short introduction to SPLs and their evolution.

\subsection{Basic Concepts}

A \emph{software product line} (SPL) is ``a set of software-intensive systems sharing a common, managed set of features that satisfy the specific needs of a particular market segment or mission and that are developed from a common set of core assets in a prescribed way''~\cite{Clements2002}.
With an SPL, customers' requirements are coded in a \emph{configuration} that is used to automatically derive a concrete \emph{product}, which is thus tailored to the customers' needs.
SPL engineering comprises all activities involved in conceiving, developing, and maintaining an SPL.

One particular activity of interest in SPL engineering is \emph{variability modeling}, in which domain engineers analyze, determine, and model an SPL's variability.
The most common way to do so is by employing \emph{features}, which represent functionalities that can be either present or absent in a product derived from the SPL~\cite{Apel2013a, Berger2013}.
Variability modeling plays an important role in SPL evolution, where features may be added, removed or modified over the course of an SPL's lifetime.

Typically, features are associated with assets, such as code or requirements, which then play a role in deriving a product~\cite{Apel2013a}.
For example, the Linux kernel (which can be considered an SPL) includes a \emph{Networking} feature, which, when selected in a configuration, adds code to the derived Linux kernel that enables networking scenarios.
Thus, the evolution of a feature's assets (i.e., changes to the code) must also be considered for SPL evolution.

\subsection{Maintenance and Evolution}

The terms \emph{maintenance} and \emph{evolution}, with regard to software development in general, are not clearly distinguished in the literature~\cite{bennett2000software, von1995program}.
\emph{Software maintenance}, as defined by IEEE Standard 1219~\cite{IEEE1219}, refers to  the ``modification of a software product after delivery to correct faults, to improve performance or other attributes, or to adapt the product to a modified environment''.
In this definition, the \emph{post-delivery} aspect is crucial; that is, software is maintained by the vendor's ongoing provision of support, bug fixes etc.
McGregor~\cite{McGregor2003} defines \emph{software evolution} as the ``accumulated effects of change over time'', where ``forces drive the change in a certain direction at a certain point in time, and whether those forces are anticipated or controlled is uncertain''.
Thus, software evolution relates closely to its maintenance.
However, maintaining a software system usually only involves preserving and improving its existing functionality; whereas evolution can also comprise major changes to the software, such as the introduction of new functionality~\cite{bennett2000software}.

The notion of maintenance and evolution is generalizable to entire SPLs~\cite{McGregor2003, Marques2019, svahnberg1999evolution, Botterweck2014}, which has even solicited a dedicated international \emph{VariVolution} workshop~\cite{Linsbauer:2018:IWV:3233027.3241372}.
In particular, new methods are required to address evolution scenarios that explicitly address an SPL's variability, and to provide safety guarantees about such evolutions (discussed in the following section).
Continuous neglection of evolutionary issues can lead to an SPL's \emph{erosion}, that is, a deviation of the variability model to a degree where desired key properties of the SPL no longer hold~\cite{johnsson2000quantifying, Marques2019}.

\section{Methodology}
\label{sec:method}

The results of this survey build on a 2018 study by~\citet{Rabiser2018}, where a random subset (140) of all (593) papers published at the \emph{International Systems and Software Product Line Conference} (SPLC) has been analyzed, according to whether their authors are from academia or industry.
The major research question of their work was whether phases, activities, and topics addressed in academic and industry SPL research align with each other.
They find that, in spite of the common assumption that academic SPL research has been drifting apart from the industry in past years, there is little evidence to support this claim.
In their study, they also classify the considered papers according to different SPL phases, activities, and topics.
For details on the methodology and classification, we refer to the original study~\cite{Rabiser2018}.
Out of 140 papers, 20 have been classified under the (single) activity \emph{maintenance/evolution}.
We take these 20 papers as the basis of our survey, as they are well-distributed over 20 years of SPLC and include a representative number of papers from academia (12) as well as industry (8).
Naturally, this selection of papers does not cover the full extent of papers on SPL evolution published at SPLC; nonetheless, it is a representative subset because of the uniformly distributed selection process.

\section{Survey}
\label{sec:results}

In this section, we present the main results of our survey.
We split our results according to the primary affiliation of the authors with academia or industry as classified originally by~\citet{Rabiser2018}.

\subsection{Insights from Academia}
\label{sec:academia}

\begin{table}
	\caption{Reviewed papers classified as academic research.}
	\label{tab:academia}
	
	\begin{tabular}{rll} 
		\toprule
		Year & Authors & Topic \\
		\midrule
		1998	& \citet{Weiderman1998} & Migration towards SPL \\
		2001	& \citet{Svahnberg2001} & Industry collaboration \\
		2008	& \citet{Dhungana2008} & Industry collaboration \\
		&  & Evolution operators \\
		2012	& \citet{Seidl2012} & Evolution operators \\
		2012	& \citet{DeOliveira2012} & Change impact analysis \\
		2012	& \citet{Rubin2012}	 & Change impact analysis \\
		2013	& \citet{Linsbauer2013} & Change impact analysis \\
		2014	& \citet{Quinton2014} & Evolution operators \\
		2015	& \citet{Teixeira2015} & Change impact analysis \\
		2016	& \citet{Sampaio2016} & Change impact analysis \\
		&  & Evolution operators \\
		\bottomrule
	\end{tabular}
\end{table}

Of the reviewed papers, \citeauthor{Rabiser2018} classify twelve as academic research, that is, the majority of authors are from an academic context (e.g., universities and research institutes).
We omit two papers from our review, a systematic mapping study and an approach for synthesizing attributed feature models, both of which do not address evolution directly and, thus, seem to be misclassified~\cite{Marimuthu2017a, Becan2015}.
In \autoref{tab:academia}, we summarize the academic papers reviewed in this section.
We categorize the papers with regard to the topics addressed, and highlight how they can be used to solve specific evolutionary problems.
We discuss several individual papers in detail, but we also group similar papers according to their topic (cf.~\autoref{tab:academia}) where appropriate.
We begin with the oldest paper of our reviewed academic papers, which motivates the problems encountered in SPL maintenance and evolution.

\paragraph{Migration towards an SPL}

\citet{Weiderman1998} investigated the question whether it is feasible to extract SPLs from existing stovepipe systems (i.e., legacy systems with limited focus and functionality that do not allow to share data easily).
Although classified as \emph{maintenance/evolution} by \citeauthor{Rabiser2018}, this paper rather focuses on how to evolve \emph{towards} an SPL.
Nonetheless, it is a first effort in this area with some key insights:
First, developers and researchers tend to focus on building new systems, rather than evolving legacy systems.
In particular, the migration of stovepipe systems towards a more structured SPL approach was largely unrecognized at the time.
Second, they noticed that the rising importance of the Internet in the late '90s began to enable interconnection of such systems, which further unraveled their weakness in sharing data with other systems.
Third, they recognized that for successful evolution, a high-level understanding of the evolved system is crucial (i.e., software modules may be treated as black boxes with defined interfaces); an insight that also applies to continuous evolution of an SPL.

\paragraph{Evolution operators}

A common pattern found in the reviewed academic papers is the identification of so-called \emph{evolution operators} (also termed templates, edits or simply evolutions)~\cite{Sampaio2016, Seidl2012, Quinton2014, Botterweck2010b, Dhungana2008}.
Such evolution operators specify different situations that can arise in SPL evolution scenarios, usually involving changes in an SPL's variability model.
Commonly specified evolution operators include \emph{adding, removing, splitting} and \emph{merging features} and \emph{asset changes} (e.g., modification of a feature's source code).
It is surprising that such operators are frequently redefined, which is most likely because they are tailored to the specific paper's research question.
For example, \citet{Seidl2012} focus on maintaining the integrity of an SPL's mapping from features to their implementation artifacts, so they adapt their evolution operators accordingly.
A single analysis and formalization of evolution operators, however, would be beneficial, as it would provide a common, compatible ground for different approaches and facilitate the implementation of tool support.
It should be noted that evolution operators are not the only possibility for encoding change:
Difference or \emph{delta models}~\cite{schaefer2010delta, Botterweck2014} serve a similar purpose, although they are not discussed in the reviewed selection of papers.

\paragraph{Change impact analyses}

A \emph{change impact analysis} is used to predict which assets a specific change will affect~\cite{McGregor2003}.
Thus, it can support evolution endeavors, as engineers can confirm how a change will affect the SPL as a whole.
Several of the reviewed papers relate to this concept.

First, \citet{Sampaio2016} present an approach for the \emph{partially safe evolution} of SPLs.
To understand this concept, consider \emph{fully safe evolution}~\cite{Neves:2015:SET:2794082.2794113}:
In this notion, an evolution is \emph{fully safe} if, for every product of the (unaltered) SPL, a product with compatible behavior exists in the SPL even after the change has been applied.
Thus, no existing users are inadvertently affected by any performed change.
However, this notion of fully safe evolution is very strict:
For example, suppose there is a (breaking) change to the Linux network stack.
Obviously, this affects all existing users that have chosen to include the networking stack into their kernel; thus, it is not considered a fully safe evolution.
However, users that deselected the networking stack are guaranteed to not be affected.
\citeauthor{Sampaio2016} formalize this concept as a \emph{partially safe evolution}, that is, an evolution that may affect some, but not all products of an SPL.
This approach has more practical applications than fully safe evolution:
For instance, this concept can help to maintain support for existing configurations by only allowing evolution that is partially safe with regard to these configurations.

\citet{Teixeira2015} further extend the safe evolution concept to account for \emph{multi product lines}, that is, a composition of product lines, such as the Linux kernel paired with the Ubuntu distributions.
They describe several concrete bugs that arise from unwanted interactions between evolutionary changes in Linux and Ubuntu and could have been avoided with suitable guidelines.
They present and formalize such guidelines, however, it remains to be seen how these concepts can be implemented in practice to aid in an industrial context.

\citet{Linsbauer2013} take a slightly different approach:
Instead of considering the safety of an evolutionary change, they focus on recovering traceability between features and code.
This is not always a trivial problem; in fact, feature interactions and implementation techniques can obscure the consequences of a change in the code.
Thus, they propose a \emph{traceability mining} approach to recover this information, with promising results (correct identification in 99\% of cases) on several non-trivial case studies.
Note that this approach is primarily aimed at the maintenance of an SPL, that is, preventing new bugs from being introduced into the product line.

In the same vein, \citet{DeOliveira2012} present an approach to analyze the \emph{bug prevalence} in SPLs with a \emph{product genealogy}, that is, a tree that reflects how products are evolved and derived from each other.
In this context, \emph{bug prevalence} refers to the number of products affected by a particular bug, and the genealogy tree may be used to determine which and how many products are \emph{infected} by this bug, making this approach interesting in the context of SPL maintenance.
However, their approach has not been evaluated in an industrial context yet.

\citet{Rubin2012} address the problem of maintaining forked product variants (also known as \emph{clone-and-own}).
As part of their approach, they propose the \emph{product line changeset dependency model} (PL-CDM), which is a meta-model that encodes information about an SPL, its products, features, and relationships between those features.
This PL-CDM may be integrated with software configuration management (SCM) systems; for example, to notify the developer of a certain product that another developer's changes affect this product.
Thus, their approach can be used to assist in change impact analysis when maintaining forked product variants.
However, their approach is visionary and has, to the best of our knowledge, neither been implemented nor evaluated.

\paragraph{Industry collaborations}

The last two papers discussed in this section each address a collaboration with an industry partner.
Nevertheless, we include them in this section, as \citet{Rabiser2018} have classified them as academic research according to their authors' affiliations; that is, they are written from an academic point of view.

\citet{Dhungana2008} report about their experience with SPLs at Siemens VAI.
They found that motivating factors for SPL evolution include new customer requirements, technological changes, and internal improvements of the SPL's assets.
Further, they identified several issues their industry partner faced:
First, the knowledge about variability in an SPL may be scattered over multiple teams, which hampers efficient collaboration on evolutionary changes~\cite{Yi2012}.
Second, different parts of the system may evolve at different speeds.
Third, meta-models (i.e., models that determine the structure of an actual variability model) may also co-evolve in parallel with variability models.
Their proposed solution involves dividing variability models into fragments, which can be evolved independently and finally merged together.
Their approach has been used and refined successfully at Siemens VAI using real-world models.

\citet{Svahnberg2001} describe an industrial example in the automatic guided vehicle domain, where the company Danaher Motion Särö AB migrated from their previous, hardware-centered generation of an SPL to a new, software-centered one.
The motivation for this was that the hardware in the domain was subsequently being replaced with software solutions, which posed some new challenges the old SPL could not meet.
This represents a rather large evolutionary change, so that they had to plan their course of action carefully.
Thus, they employed an iterative process, where the old SPL is subsequently replaced by the new generation.
Notably, they place large emphasis on continuous support, as customary in their domain:
Customers' configurations should be actively maintained and supported for at least ten years.
To achieve this, the exact configurations of all customers need to be tracked, and it must be carefully investigated whether and how evolutionary changes may affect them.
For the latter, the partially safe evolution approach proposed by \citet{Sampaio2016} may be of use.

Similar to \citeauthor{Dhungana2008}, \citeauthor{Svahnberg2001} further struggled with finding the right granularity for their evolution, that is, whether to migrate a given component completely, split it into different parts or eliminate it entirely from their architecture.
However, at the time the paper was written, the migration process had only been partly concluded, so it is not known whether this large evolution eventually succeeded.
In later work, Svahnberg et al.~\cite{doi:10.1002/spe.652, Svahnberg2003} have further compiled a taxonomy of variability realization techniques based on this and other industrial collaborations, which is intended to guide the selection of a software architecture suitable for graceful evolution of SPLs.

\subsection{Insights from Industry}

\begin{table}
	\caption{Reviewed papers classified as industrial research.}
	\label{tab:industry}
	
	\begin{tabular}{rll} 
		\toprule
		Year & Authors & Topic \\
		\midrule
		2000	& \citet{Dager2000}	& Migration towards SPL \\
		2001	& \citet{VanOmmering2001} 	& Development process \\
		2002	& \citet{VanDerLinden2002}	& Project description \\
		2003	& \citet{Karhinen2003}	& Knowledge management \\
		2009	& \citet{Pech2009}	& Variability management \\
		2011	& \citet{Vierhauser2011}	& Deployment \\
		2012	& \citet{Martini2012}	& Speed and reuse \\
		2013	& \citet{Zhang2013}		& Variability erosion \\
		\bottomrule
	\end{tabular}
\end{table}

The remaining eight reviewed papers are motivated mainly from industry, that is, companies and their research departments, as classified by \citet{Rabiser2018}.
In \autoref{tab:industry}, we summarize the industrial papers reviewed in this section.
Similar to the academic papers, we categorize the reviewed papers regarding their addressed topic.
In the following, we discuss the papers individually in chronological order.
We further connect the dots between these industrial papers, and correlate them with discussed academic papers, where possible.
Similar to above, we start with an early paper that examines the SPL maintenance and evolution problem from an industrial perspective.

\paragraph{Cummins}

\citet{Dager2000} describes the development and maintenance of an SPL architecture at Cummins, a large manufacturer of diesel and gas engines.
Up until the early '90s, Cummins had a practice of starting from scratch for every new developed product.
However, management soon realized that to preserve and consolidate Cummins' market position, new ideas were needed, as the old approach was becoming uneconomical and could no longer serve the growing diversity in customer requirements.
In addition, software engineering gradually became more important for developing embedded systems, to the point that software is now as much an asset to Cummins as the actual hardware.
These factors motivated a big effort at Cummins to transition to an SPL approach, which \citeauthor{Dager2000} describes in detail; here, we focus on Cummins' experiences related to SPL maintenance and evolution.

The initial switch to an SPL architecture at Cummins went very well:
A core team was established to develop central software artifacts, which could then be reused throughout the company by the market and fuel systems teams for different products.
This concept was simple, but immediately paid off, resulting in faster time-to-market, improved portability, and more predictable product behavior for the end customers.
However, the concept began to falter under the huge amount of product requirements added onto the system, and within five years, software reuse deteriorated by more than 50\%.
The core team traced this observation back to several issues.
Among the identified problems were a lack of a proper evolution process and architecture documentation to keep the end products maintainable.
In particular, Cummins realized that adopting SPLs would not lead to optimized product reuse on its own; instead, this transition should be accompanied by institutional changes as well as new roles and responsibilities matching the SPL approach.

A new effort was made to improve the first SPL approach by addressing these problems.
The reconsideration of Cummins' business needs showed that software reusability is only one driver of profitable product development; however, neglecting other drivers would prevent Cummins from unlocking the full potential of their products.
Among these other drivers, \citeauthor{Dager2000} notes, are easy maintenance and evolution of the SPL, which had not been a focus in the initial SPL approach.
In particular, stable software components are expected not to break due to maintenance efforts, which is an example for partially safe evolution as proposed by \citet{Sampaio2016}.
Because the identified cost drivers may be opposed, Cummins needed to carefully weigh and study these factors for each developed product in the future.
According to \citeauthor{Dager2000}, this new approach can be expected to be more successful, due to increased awareness of other requirements and key drivers for successful SPL engineering---however, the paper provides no information on the outcome of this endeavor.

\paragraph{Philips}

Van Ommering~\cite{VanOmmering2001} proposes an approach for planning ahead the architecture of \emph{product populations} by combining a top-down SPL approach with a bottom-up reusable component approach.
At Philips, a product population is considered a family of product lines; for example, TV and VCR devices each form a product line, while together they form a product population.
The problem with applying SPL concepts directly to product populations is that SPLs are usually organized around a common architecture with individual variation points; a top-down design that lends itself well to products with high similarities and few differences.
However, TV and VCR devices, for example, are not represented easily in this design, as they have many differences.
In this context, reusable software components are better-suited, as they are designed to be used in any context.
This bottom-up approach, however, lacks the holistic view on the product population provided by an SPL approach, so that components can end up too generic and therefore costly to develop and maintain.

Van Ommering suggests to combine the SPL and software component approaches by contextualizing all developed software components in their usage context according to the SPL.
To this end, \citeauthor{VanOmmering2001} introduces a planning scheme to bring together component developers and product engineers, where components offered by the developers are associated with the requirements of product engineers.
Notably, the planning scheme involves a time component, which allows to plan for maintenance and evolution.
This way, evolution steps (such as adding and removing functionality) are modeled explicitly, which is an important requirement for Philips.
In particular, future evolution steps can be planned ahead, and automatic consistency checks are possible.
In more recent work, this idea has resurfaced in the form of \emph{temporal feature models}, which consider evolution a first-class dimension in variability modeling~\cite{Nieke2016}.
According to an extension paper~\cite{DBLP:conf/icse/Ommering02}, Philips successfully applied this planning scheme to four business groups for several types of TVs, which also necessitated some organizational changes, similar to what \citet{Dager2000} observed at Cummins.

\paragraph{ESAPS}

From 1999 to 2001, the \emph{Engineering Software Architectures, Processes and Platforms for System Families} (ESAPS) project was performed, which resulted in a cooperation between 22 research institutes and companies (including Philips, Nokia, Siemens, Thales, and Telvent).
Van der Linden~\cite{VanDerLinden2002} summarizes the goals of ESAPS, that is, improving both development paradigms and reuse level in software development.
The paper also describes the results of the project, where we again focus on maintenance and evolution.
Notably, a distinction of variability arose within ESAPS, namely \emph{variability-in-the-large} and \emph{variability-in-the-long-term}:
The first term refers to SPLs with many products existing at any given time (e.g., consumer products), while the second term refers to SPLs with few products at a time, but many over the course of time (e.g., professional products).
The project identified differences between these two cases, for instance, in their software architecture.
The latter, variability-in-the-long-term, usually involves evolution, as products are continuously enhanced.
The project recognized this and particular attention was given to change management (similar to change impact analysis, cf.~\autoref{sec:academia}).
Guidelines, processes, and hierarchies for change management were proposed and introduced accordingly, but not described further by \citeauthor{VanDerLinden2002}, and the detailed project report is no longer available.
The success of ESAPS led to a follow-up project, CAFÉ, in which change impact analysis should also be further investigated~\cite{van2002software}.

\paragraph{Help desk}

\citet{Karhinen2003} emphasize the importance of a well-designed software architecture in a high-quality SPL.
They argue that together with an SPL's software architecture, the needs for communication evolve as well, because the long-term evolution of an SPL can also involve organizational and personal changes (e.g., newly recruited employees).
This mirrors the experiences of \citet{Dager2000} and \citet{VanOmmering2001} described above.
Thus, they propose to install a \emph{software architecture help desk} in the organization, which is its focal point of communication regarding software architecture.
This help desk is intended to assist not only in initial development, but also in the subsequent maintenance and evolution of an SPL.
During these phases, the help desk is a point of reference for all employees:
It guides developers with the design and evolution of an SPL's architecture, for instance by giving courses and answering questions.
Thus, the help desk steers the software architecture and prevents divergence from organization-wide policies (e.g., due to local developer communities).
\citeauthor{Karhinen2003} have put this concept into practice in cooperation with two industry partners.
However, they only report early results concerning the conception phase of the developed SPLs.

\paragraph{Wikon}

\citet{Pech2009} describe how Wikon, a small company in the embedded systems domain, switched to a decision modeling approach for one of their variability-intensive software systems.
Initially, Wikon maintained a variable software system based on conditional compilation with C preprocessor directives (e.g., \texttt{\#ifdef} and \texttt{\#define}), coupled with custom header files for each product to be derived.
To manage variability, these header files had to be edited manually, which was tedious and error-prone, so that the three responsible developers spent much time only on maintenance.
An increasing number of products and variation points necessitated more explicit variability modeling, which Wikon implemented with assistance from Fraunhofer IESE.
One particular goal of this transition was to facilitate maintenance and evolution of the modeled variability.
The implementation of a decision modeling approach ensured this in two ways:
First, the usage of one common decision model ensures that all derived products (in the form of custom header files) are syntactically valid.
This prevents that an evolutionary change accidentally overlooks one of the custom header files; for example, when removing some functionality from the SPL.
Second, the decision model comprises several \emph{constraints}, which explicitly model domain knowledge that had been tacit before.
Thus, an automated system can check the internal consistency of a product; that is, whether the product actually conforms to the modeled domain.
This was helpful because only one of the three responsible developers at Wikon had the necessary domain knowledge, and the others could now rely on the automatic consistency checking.
\citeauthor{Pech2009} successfully introduced this approach:
A new product was easily derived, and according to the developers, the maintenance effort for the entire SPL after the transition was reduced to the maintenance effort for a single product before the transition.

\paragraph{Siemens}

\citet{Vierhauser2011} describe a deployment infrastructure for product line models and tools, which is used in a family of electrode control systems at Siemens VAI.
In their product line, sales documents like customer-specific offers (the products) can be automatically derived from document templates (the artifacts) to speed up the process of creating an offer and also prevent legal issues.
This product line, however, is frequently updated and must be re-deployed to all sales people, which includes updating the actual product line models, but also the surrounding tooling.
To this end, \citeauthor{Vierhauser2011} contribute a technical approach based on \emph{product line bundles}.
Notably, their approach involves setting an expiry date for outdated product line models, so that sales people (which often work offline) are forced to use the latest, updated model.
Furthermore, a mechanism is provided to migrate a product based on an outdated model to the newest model version.
However, \citeauthor{Vierhauser2011} do not report on experiences with this approach.

\paragraph{Speed and reuse}

\citet{Martini2012} analyze how systematic software reuse can be combined with an agile software development process.
They identify two aspects with profound impact on the success of an agile process, namely \emph{speed} and \emph{reuse}.
High speed is required to compete with other contenders, while high reuse has positive effects on productivity and thus profit.
In a case study with three organizations, \citeauthor{Martini2012} identify and categorize 114 factors that influence speed and reuse in different ways.
They distinguish three kinds of speed: \emph{first deployment}, \emph{replication}, and \emph{evolution speed}.
Evolution speed, which is relevant for SPL maintenance and evolution, is described as the ``time used to decide whether an incoming evolution request should be addressed or not, and the time for actually developing it''~\cite{Martini2012}.
A high evolution speed is necessary shortly after introducing a new product to roll out urgent updates.
While the product is on sale, evolution speed is also important to be able to compete on the market.
\citeauthor{Martini2012} find that the studied organizations focused first and foremost on first deployment speed (the time until the product is first released), rather than replication or evolution speed.
They also found that, when categorizing the most harmful factors regarding speed, key areas for improvement in the studied cases are communication and knowledge management.
This again reflects the findings of \citet{Dager2000}, \citet{VanOmmering2001}, and \citet{Karhinen2003} in the context of agile software development.
Unfortunately, \citeauthor{Martini2012} did not look further into evolution speed in their paper or technical report~\cite{martini2012factors}.

\paragraph{Danfoss}

In the final paper reviewed in our survey, \citet{Zhang2013} investigate the long-term evolution of a large-scale industrial SPL at Danfoss Power Electronics.
Their research is motivated by the observation that uncontrolled evolution (in particular, adding new variability and, thus, variants) can lead to a reduction of productivity, which manifests in the erosion of an SPL (i.e., artifacts that get much more complex and inconsistent over time).
Based on their industrial experiences, they claim that there are no proper tools and methods for detecting, removing, and preventing such erosion---thus, organizations tend to create a new SPL instead, which is expensive and contradicts, for instance, the ten-year support requirement of \citet{Svahnberg2001}.
Thus, \citeauthor{Zhang2013} aim to identify tactics for detecting and forecasting SPL erosion to sustain the productivity of an evolving SPL even under economic constraints.

For their actual analysis, \citeauthor{Zhang2013} consider 31 versions (each 3.6 MLOC on average) of an SPL of frequency converters, covering a period of four years of maintenance and evolution history.
They identify several metrics for measuring SPL erosion based on the conditional compilation technique in C (i.e., \texttt{\#ifdef} annotations), which they mined from the SPL's source code with the \emph{VITAL} tool suite.
Among the examined metrics are the number of features, variability nesting, tangling, and scattering, as well as the complexity of variable source code files.
\citeauthor{Zhang2013} find that \texttt{\#ifdef} nesting, tangling (files that contain code for many different features), and scattering (features that are referred to from many different files) lead to a high maintenance cost if the affected features are likely to evolve.
They propose to solve these problems by refactoring such code, for example, towards an aspect-oriented approach.
By calculating metrics for all versions with VITAL, \citeauthor{Zhang2013} also determine that nesting, tangling, and scattering are often introduced gradually during the evolution of the SPL.
Consequently, an erosion trend can be calculated from the history, so that erosion metrics for future SPL versions can be predicted.
Thus, developers are able to identify potential erosion hotspots in advance and take measures to avert erosion.
However, the approach of \citeauthor{Zhang2013} is only applicable in the context of the conditional compilation technique.

\section{Discussion}
\label{sec:discussion}

In this section, we discuss and compare the results from the surveyed papers. We focus on the relationship between academic and industrial papers, as well as key insights and possible future directions for SPL evolution research.

\paragraph{Academia and industry}

\begin{figure}
	\includegraphics[width=0.7\linewidth]{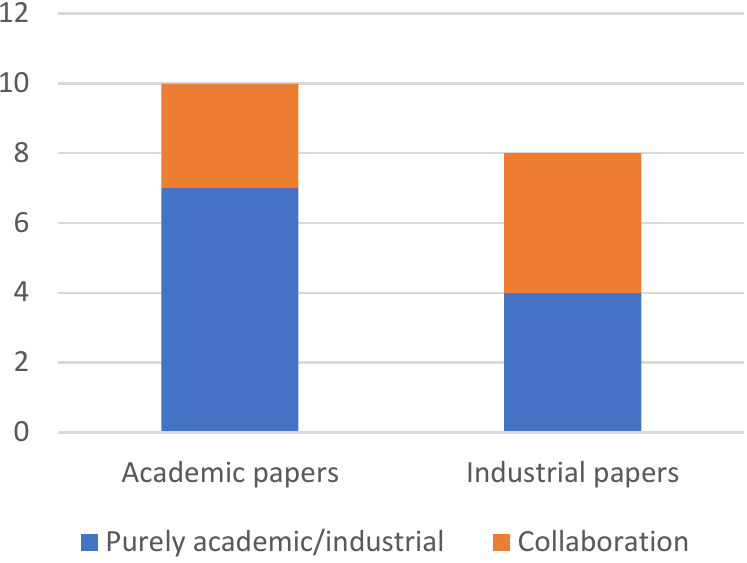}
	\caption{Classification of academic and industrial papers.}
	\label{fig:classification}
\end{figure}

In many of the reviewed academic papers, there is a clear focus on concepts, methods, and technical advancements for facilitating SPL evolution, such as the definition and study of evolution operators and change impact analyses.
On the other hand, many papers from an industrial context (unsurprisingly) tend to focus on the creation of value, economic and organizational constraints as well as human factors (such as communication and knowledge management).
We investigate this in more detail in \autoref{fig:classification}, where we show the classification of all reviewed papers into academic and industrial as performed by \citet{Rabiser2018}.
In addition, we include our assessment of this classification's accuracy.
Of ten reviewed academic papers, seven were written exclusively by authors from academia~\cite{Weiderman1998, Seidl2012, Rubin2012, Linsbauer2013, Quinton2014, Teixeira2015, Sampaio2016}.
These papers contribute novel concepts and ideas for SPL evolution and have no particular connection to industry.
Similarly, four of eight industrial papers are exclusively from an industrial context, often written by a single author---these papers mostly share experiences and measures taken by organizations to solve evolutionary issues~\cite{Dager2000, VanOmmering2001, VanDerLinden2002, Karhinen2003}.
The remaining seven papers, however, are not as easily classified, as they are collaborations between industry and academia:
Six papers (four classified as industrial, two as academic) were mostly written by authors from academia, but address industrial motivations, examples or experiences~\cite{Svahnberg2001, Dhungana2008, Pech2009, Vierhauser2011, Martini2012, Zhang2013}; also, a single paper (classified as academic) has a purely conceptual contribution although written by authors from industry~\cite{DeOliveira2012}.

From these results and the survey of \citet{RabiserMaterial} (cf.~\autoref{fig:interests}) we infer that both academia and industry are clearly aware of the importance of SPL maintenance and evolution.
Further, there is a fair amount of collaboration between academia and industry in our subset of reviewed papers, to a degree that the binary distinction between academic and industrial papers made by \citet{Rabiser2018} is only partially applicable.

\paragraph{Key insights}

Finally, we share some key insights from our survey that shed some light on specific challenges and possible future work in the field of SPL maintenance and evolution.

\begin{itemize}
	\item Successful long-term SPL evolution and maintenance is a driver of profitable product development and necessitates organizational changes (e.g., new roles and responsibilities)~\cite{Dager2000, VanOmmering2001, Karhinen2003, Martini2012}.
	Particular areas of improvement include communication and knowledge management~\cite{Karhinen2003, Martini2012}.
	
	\item Safe evolution of SPLs is requested by the industry to satisfy stability and support requirements~\cite{Dager2000, Svahnberg2001, VanDerLinden2002} and has been investigated by academia~\cite{Sampaio2016, Teixeira2015, DeOliveira2012}, although these techniques are not widely adopted by industry yet.
	More change impact analyses and traceability approaches have been proposed by academia~\cite{Linsbauer2013, Rubin2012}---in principle, this line of work has been acknowledged as important by the industry, although not adopted yet~\cite{VanDerLinden2002, Zhang2013}.
	
	\item There is a lack of publicly available industrial-sized case studies and long-term evolution histories~\cite{Marques2019}.
	Further, experience reports typically only report early-stage results~\cite{Svahnberg2001, Dager2000, Karhinen2003, Vierhauser2011}; only one of the reviewed papers performs an empirical investigation on an evolving SPL~\cite{Zhang2013}.
	This may be due to the increased effort of a long-term study or because industry partners are reluctant to share their SPLs.
	
	\item SPL erosion is a known, but rarely investigated problem~\cite{Marques2019}. The term is not clearly defined, although it usually relates to a reduction in productivity as software is aging~\cite{Dager2000, Zhang2013}.
	Among others, erosion is promoted by a lack of evolution planning~\cite{VanOmmering2001, Zhang2013, Nieke2016}.
	
	\item SPL evolution research tends to focus on adding and maintaining variability, while removal of variability is rarely considered~\cite{Berger2014, Marques2019}.
	However, only adding variability can lead to a gradual increase of complexity and, thus, to SPL erosion~\cite{Zhang2013}.
	Instead, variation points should be continuously reconsidered and removed as soon as they become obsolete to prevent erosion in the first place.
	Awareness for this issue still has to be raised~\cite{Zhang2013, Marques2019}---for instance, future research may investigate whether the concept of (partially) safe evolutions is reconcilable with the notion of removing variability.
	
	\item The adoption of more advanced tooling for variability management (such as \emph{pure::variants}) can lead to a significant acceleration of code size and growth in variability~\cite{Zhang2013}.
	On the one hand, such improvements in tooling allow more efficient development and, thus, a faster time-to-market.
	However, this also promotes SPL erosion in the long run and must therefore be carefully considered~\cite{Zhang2013}.
\end{itemize}

\section{Threats to Validity}
\label{sec:validity}

To discuss threats to the validity of our survey, we distinguish internal and external validity~\cite{Wohlin2003, campbell2015experimental}.
To ensure internal validity, we considered a random subset of papers published at SPLC (140 of 593 total), which has been studied by \citet{Rabiser2018}.
In said study, seven researchers (each with more than 15 years of experience in the field) classified 20 papers as concerning SPL maintenance and evolution.
The identified papers are well-distributed over 20 years of SPLC and are motivated from academia (12) as well as industry (8), illustrating several key issues of the field.
Thus, the conclusions we draw from the surveyed papers can be considered sound in the context of the studied subset of 140 SPLC papers.

However, our survey suffers from a lack of external validity, as we only consider papers published at a single conference (SPLC).
Further, although the random selection of studied SPLC papers may paint a representative picture of the state of SPL evolution, key contributions to the field could have been overlooked.
Thus, our survey is by no means comprehensive, but rather an in-depth insight into specific issues of SPL maintenance and evolution.

\section{Conclusion}
\label{sec:conclusion}

In this technical report, we surveyed and discussed selected papers about SPL maintenance and evolution that were published at SPLC over a period of 20 years.
We selected the papers according to an existing classification by \citet{Rabiser2018} and make use of their differentiation between academic and industrial papers.
We further extended the work of \citeauthor{Rabiser2018} with an in-depth discussion of evolutionary issues covered in the reviewed papers.
We found that 7 of 18 reviewed papers can be considered collaborations between industry and academia, which suggests that both are rather in line with each other.

Although both practitioners and researchers are aware of the importance of SPL evolution, our survey suggests that several problems found in industrial applications are still unsolved.
In particular, future work should look further into the erosion of SPLs, its long-term effects on an SPL, and how to remove or even prevent it.
Further, the field would benefit from a larger number of publicly available case studies originating in industry.

\balance
\bibliographystyle{ACM-Reference-Format}
\bibliography{abbreviations,IEEEfull,MYfull,paper,thesis,studconf}

\end{document}